\documentclass[aps,prxquantum,reprint,superscriptaddress,longbibliography]{revtex4-2}
\usepackage{amsmath}
\usepackage{amssymb}
\usepackage{graphicx}
\usepackage{physics}
\usepackage{tikz}
\usetikzlibrary{quantikz}
\makeatletter
\newcounter{algorithm}
\def\fps@algorithm{tbp}
\def\ftype@algorithm{8}
\def\ext@algorithm{loa}
\def\fnum@algorithm{Algorithm~\thealgorithm}
\usepackage{algorithm}
\usepackage{algpseudocode}

\newlength{\qctarget}
\newlength{\qcunit}
\newsavebox{\qcbox}
\newcommand{\qcsolve}[2]{%
  \sbox{\qcbox}{#2}%
  \setlength{\qcunit}{\dimexpr\qcunit + (\qctarget - \wd\qcbox) / #1\relax}}
\newcommand{\fitcircuit}[2]{%
  \setlength{\qcunit}{\dimexpr\qctarget / #1\relax}%
  \qcsolve{#1}{#2}\qcsolve{#1}{#2}%
  \sbox{\qcbox}{#2}%
  \typeout{fitcircuit: unit=\the\qcunit, width=\the\wd\qcbox, target=\the\qctarget}%
  \resizebox{\columnwidth}{!}{\usebox{\qcbox}}}
\usepackage{multirow}
\usepackage[colorlinks=true,allcolors=blue]{hyperref}

\begin{document}

\title{Parallel and Distributed Fermionic Simulation via Dynamic Encoding}

\author{Michael Williams de la Bastida}
\email{michael.williams.20@ucl.ac.uk}
\affiliation{Centre for Computational Science, Department of Chemistry, UCL}
\author{Peter V. Coveney}
\affiliation{Centre for Computational Science, Department of Chemistry, UCL}

\begin{abstract}
We demonstrate a simple and efficient method to parallelize and distribute Trotterized Hamiltonian simulation of fermionic systems across multiple QPUs. Using combinatorial covering designs to define a fixed set of fermion-qubit encodings, independent of system size, we demonstrate communication cost scaling as $\mathcal{O}(M q^3 r)$ for a system of $M$ fermionic modes and Trotter number $r$. This improves on the static encoding bound for $q$ QPUs, $\mathcal{O}(M^4 q r)$. In terms of QPU register size $R$, at most $R^{1/3}$ QPUs are able to parallelize at most $R^{4/3}$ fermionic modes. We compare this approach to dynamic encoding using a randomized method, Pauli-weight based optimization and hypergraph partitioning. Applying these to the Hamiltonians of a range of molecular systems split across two and three QPUs, we find the combinatorial covering approach results in significantly reduced communication cost.
\end{abstract}

\maketitle

\section{Introduction}
Simulations of chemistry and materials are expected to be among the first computational tasks to benefit from quantum advantage. Current quantum processing units (QPUs) still suffer from a lack of physical qubits, restricting the size of inputs, and sensitivity to noise, limiting the number of operations which can be used in each computation.

As individual QPUs become larger, control and noise mitigation become more challenging. One possible approach to increasing the size of admissible circuits is to distribute portions of the computation between multiple QPUs.~\cite{COSMACommunicationawareOptimization2026,LinearComplexityFermionic2026,DistributionComplexityElectronic2026,DistributedQuantumComputing2022} Explorations of this to date suggest that inter-QPU entangling operations will present a new bottleneck, as gate fidelities are significantly lower than those carried out on-device.~\cite{DistributedQuantumComputing2025,DistributedQuantumComputing2022} Distributed computation will require specialized circuit compilation and error correction techniques which aim to minimize these costs.~\cite{DistributedQuantumSimulation2024a, ReviewDistributedQuantum2025} While quantum circuit compilers largely employ algorithm-agnostic optimization methods, the inclusion of algorithm-specific properties and invariants in circuit building can provide further benefit.

One such opportunity arises when simulating fermionic Hamiltonians which describe properties of chemistry and materials. To represent fermionic operators on a digital quantum computer, a fermion-qubit encoding must be selected. There are many valid encodings which produce quantum circuits of vastly different quality depending on the target Hamiltonian and processing hardware.~\cite{TernaryTreeTransformations2024,UeberPaulischeAequivalenzverbot1928,SuperfastEncodingsFermionic2019,OptimalFermiontoqubitMapping2020} Further, many optimization methods exist to select between and within classes of encoding. Some of these target features of the processor~\cite{BonsaiAlgorithmGrow2023, TreespilationArchitectureStateOptimised2024} while others target the operators to be encoded~\cite{HATTHamiltonianAdaptive2025,OptimisedFermionQubitEncodings2026,OptimalFermionqubitMappings2025,HuffmanCodebasedTernaryTree2025}. Naturally, considering features of both operators and hardware leads to workflows which resemble fermionic compilers more than straightforward encodings.~\cite{TreespilationArchitectureStateOptimised2024,FermihedralOptimalCompilation2024a} By incorporating additional fermionic swap operators into a circuit, a \emph{dynamic} encoding scheme can be defined with properties not available to a single \emph{static} encoding.~\cite{FastSimulationFermions2025,LinearComplexityFermionic2026}
 
Several recent publications have demonstrated fermionic compilation methods to improve the parallelism and distribution of simulation methods for chemistry and materials science.~\cite{COSMACommunicationawareOptimization2026,LinearComplexityFermionic2026,DistributionComplexityElectronic2026} These have been limited to specific problem formulations split between a pair of QPUs,~\cite{DistributionComplexityElectronic2026} optimization over a single fermion-qubit encoding~\cite{COSMACommunicationawareOptimization2026} and on-device parallelism using dynamic encoding~\cite{LinearComplexityFermionic2026,QuantumSimulationElectronic2018}.

In this work we demonstrate a simple and efficient method for parallelizing Trotterization of fermionic Hamiltonians with communication cost scaling linearly in the number of fermionic modes and the Trotter number. Section~\ref{sec:Trotter} begins with an overview of Trotterization-based Hamiltonian simulation. Encodings of fermionic operators onto qubits are discussed in Section~\ref{sec:encoding}. This is followed in Section~\ref{sec:coverings} by a brief introduction to combinatorial coverings.
% , with their application to fermionic circuits in Section~\ref{sec:circuit}.
We present our method for distributed fermionic Trotterization in Section~\ref{sec:pft}. Results follow in Section~\ref{sec:communication-cost}, with discussion relating our work to existing methods in Section~\ref{sec:discussion}.

\section{Trotterization of the Electronic Structure Hamiltonian}\label{sec:Trotter}
Many properties of chemistry and materials can be described by the electronic structure Hamiltonian (ESH) under the Born-Oppenheimer approximation. In second quantization this is composed of fermionic creation $a^\dagger$ and annihilation $a$ operators, together with coefficients $c_\mathrm{ij} \in \mathbb{R}$ for the one-electron term and $c_\mathrm{ijkl} \in \mathbb{R}$ for the interaction term:
\begin{equation}
  H = \sum_\mathrm{ij}^M c_\mathrm{ij} a^\dagger_i a_j + \sum_\mathrm{ijkl}^M c_\mathrm{ijkl} a^\dagger_i a^\dagger_j a_k a_l
  \label{eq:esh}
\end{equation}
To simulate time evolution under a Hamiltonian, we must implement a quantum circuit equivalent to the exponential of this
\begin{equation}
  \psi(t) = e^{-i H t} \psi(0)
\end{equation}
For Hamiltonians composed of a sum of terms, such as in Eq.~\eqref{eq:esh}, the Lie-Trotter formula can be applied to transform this exponential of a sum into a product of exponentials. Approximations to this are given by the Trotter-Suzuki method, or \emph{Trotterization}
\begin{equation}
  \exp(-i H t) = \exp\Big(-i \sum_j h_j t\Big) \approx \Big(\prod_{j} e^{-i h_j t / r}\Big)^r
  \label{eq:first-order-Trotter}
\end{equation}
where $r$ is the \emph{Trotter number}, the number of repetitions of the operators, and $h_j \subseteq H$ are disjoint subsets of Hamiltonian terms. Trotterization to first order is simply an application of each term sequentially. In the limit $r \rightarrow \infty$ the method is exact, but lower Trotter number results in significant error.~\cite{TheoryTrotterError2021} Trotter number therefore presents a trade-off between simulation accuracy and the number of quantum operations used. Note that the order of terms in a circuit is not restricted, owing to the set of Hamiltonian terms being unordered. This flexibility offers an opportunity to optimize the ordering of terms in a circuit, such that a shorter circuit is compiled.~\cite{OrderingMattersStructure2025} 
Trotterization of the ESH can be carried out directly on the fermionic representation of the Hamiltonian, however to implement a circuit on a gate-based QPU a fermion to qubit encoding is needed.

\section{Fermion-Qubit Encodings}\label{sec:encoding}
An encoding $\mathcal{E}: \mathcal{F}^M_- \to \mathcal{H}^{Q}$  from the anti-symmetric Fock space of fermions to the Hilbert space of qubits must preserve the algebra of encoded operators and eigenvalues of the Hamiltonian, but there is otherwise a large degree of flexibility in exactly how this is done. The most straightforward and commonly used encoding is the Jordan-Wigner (JW) encoding.~\cite{UeberPaulischeAequivalenzverbot1928}

Under the Jordan-Wigner encoding, mode occupations are mapped directly to the spins of qubits, with fermionic exchange anti-symmetry enforced by strings of Pauli-Z operators. This results in individual fermionic operators requiring a number of Pauli operators which grows linearly in system size. However, for fermionic Hamiltonian terms of even degree, these Z operators cancel below the lower qubit index, resulting in favorable localization and parallelizability compared to the related ternary tree class of encodings~\cite{OptimalFermiontoqubitMapping2020,OptimisedFermionQubitEncodings2026}. Owing to its simplicity JW also serves as the basis for many more encoding approaches which aim to target specific QPUs or algorithms~\cite{JordanWignerMappingNonorthogonal2025,LocalJordanWignerTransformations2024,SymmetricJordanWignerTransformation2021,HigherdimensionalJordanWignerTransformation2022,UnifiedFrameworkTransformations2022,TernaryTreeTransformations2024,DiscoveringOptimalFermionqubit2023,LowdepthFermionRouting2025}. In the following, we make use of the JW encoding to parallelize the ESH but our method does not require that Hamiltonians are simulated using it.

The first step of the JW encoding is to transform fermionic operators to Majorana operators. For a system of $M$ fermionic modes, there are $2M$ Majorana operators, $\gamma_{2j} = 1/\sqrt{2}\,(a_j + a_j^\dagger)$ and $\gamma_{2j+1} = -i/\sqrt{2}\,(a_j - a_j^\dagger)$. Each Majorana operator then maps to a string of Pauli operators of the form:
\begin{equation}
  \gamma_{2j} = \frac{1}{\sqrt{2}}\Big(\bigotimes_{l<{\pi(j)}} Z_l\Big) \otimes X_{\pi(j)}
  \label{eq:m-even}
\end{equation}
\begin{equation}
  \gamma_{2j+1} = \frac{1}{\sqrt{2}}\Big(\bigotimes_{l<{\pi(j)}} Z_l\Big) \otimes Y_{\pi(j)}
  \label{eq:m-odd}
\end{equation}
As the number of fermionic modes and qubits are equal under JW, the enumeration of mode indices to qubit indices $j \rightarrow \pi(j)$ is a permutation of natural numbers $[0,M)$.~\cite{DiscoveringOptimalFermionqubit2023} In the standard, \emph{naive}, form of JW $j = \pi(j)$. 

% For Hamiltonians describing real chemical systems, some terms of the ESH in Eq.~\eqref{eq:esh} have coefficients of zero. It is therefore possible to optimize over this enumeration scheme to reduce the number of Pauli operators needed to encode only the non-zero terms.~\cite{OptimisedFermionQubitEncodings2026,CliffordCircuitBased2025a,RandomizedSubsystemDescent2026}

\subsection{Qubit Support of Fermionic Operators}\label{sec:support}
The support $S(P)$ of a Pauli-string $P$ is the set of qubits for which it acts non-trivially. Correspondingly the support $S_{\mathcal{E}}(\gamma)$ of a Majorana operator $\gamma$ under encoding $\mathcal{E}$, is the set of qubits for which the Pauli-string representation of that operator acts non-trivially. Using this, we can divide a Hamiltonian in terms of Majorana operators $H_\gamma$ into subsets according to the support of its terms.

\begin{figure}[h!]
  \includegraphics[width=0.8\columnwidth]{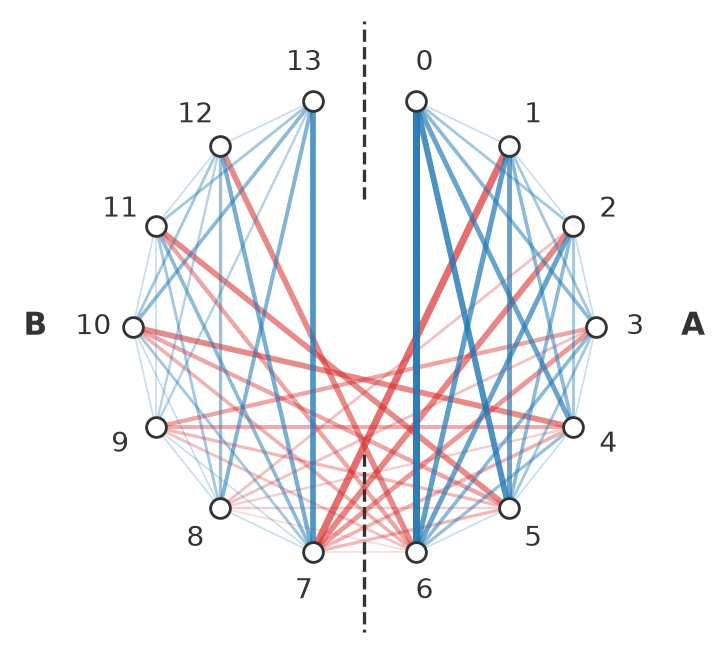}
  \caption{Support partition of operators in the naive JW-encoded water molecule in STO-3G basis. Lines show the upper and lower qubit indices for terms with support on up to half of the qubits, with thickness representing the number of terms. Lines in red cross the partition and are therefore not parallelized by the naive JW encoding.}
  \label{fig:water-partition}
\end{figure}

Given a qubit register $Q$, let $A, B$ be subsets of $Q$ such that: $A \cap B = \emptyset$ and $A \cup B = Q$. Under encoding $\mathcal{E}$, terms of $H_\gamma$ can be divided into:
\begin{equation}
  H^A = \{h \in H_\gamma \mid S_\mathcal{E}(h) \subseteq A\}
\end{equation}
\begin{equation}
  H^B = \{h \in H_\gamma \mid S_\mathcal{E}(h) \subseteq B\}
\end{equation}
\begin{equation}
  H^\mathrm{AB} = H_\gamma \setminus (H^A \cup H^B)
\end{equation}

By definition, $H^A$ and $H^B$ can be simulated on parallel QPUs as they act on disjoint sets of qubits. The terms of $H^\mathrm{AB}$ cannot be simulated in parallel with encoding $\mathcal{E}$, but we can transform this encoding into another in which additional terms of $H^\mathrm{AB}$ are parallelizable.

In the standard JW encoding, any system with an even number of modes can be partitioned in two by splitting the encoding at its midpoint. This gives two equally sized subsets of contiguous qubits with indices over the ranges $[0, M/2)$ and $[M/2,M)$. Figure \ref{fig:water-partition} shows a partition of terms in the JW encoded Hamiltonian for the water molecule in the minimal STO-3G basis. Likewise, for $q$ QPUs, the encoding can be partitioned into $q$ subsets of qubits. Note that these indices relate to virtual qubits and can be re-labelled to fit the physical qubit layout of a QPU without changing the structure of the encoding.~\cite{DiscoveringOptimalFermionqubit2023,LinearComplexityFermionic2026,OptimisedFermionQubitEncodings2026}

\subsection{Transformations Between Encodings}
The JW encoding is a special case of the broader class of Majorana-string encodings. In general, it is possible to map from one of these encodings to another with a unitary operator.~\cite{CliffordCircuitBased2025a} For two encodings $\mathcal{E}^0$ and $\mathcal{E}^1$ acting on a fermionic operator $U_f$, we can define the unitary $V_{01}$ which maps $\mathcal{E}^1 \rightarrow \mathcal{E}^0$ by conjugation,
\begin{equation}
  \mathcal{E}^0(U_f) = V_{01}\mathcal{E}^1(U_f)V^{\dagger}_{01}
  \label{eq:encoding-transform}
\end{equation}
For the ternary tree class of encodings, to which JW belongs, maps can be achieved with only a Clifford operator $C_{01}$,
\begin{equation}
  \mathcal{E}^0(U_f) = C_{01}\mathcal{E}^1(U_f)C^{\dagger}_{01}
  \label{eq:tree-transform}
\end{equation}
with controlled-NOT depth upper bounded by $2M-2$, where $M$ is the number of fermionic modes.~\cite{CliffordCircuitBased2025a}
By fixing the underlying structure of the encoding used, such that each circuit chunk uses a locally optimal enumeration of the same encoding, the boundary operator is the fermionic swap network $\mathrm{fSWAP}_{ij}:\pi^i \to \pi^j$ which implements the change of enumeration from one encoding to the next:
\begin{equation}
  \mathcal{E}^0(U_f) = \mathrm{fSWAP}_{01}\mathcal{E}^1(U_f)\mathrm{fSWAP}^{\dagger}_{01}
  \label{eq:fswap-transform}
\end{equation}
Fermionic swap gates under the Jordan-Wigner encoding are commonly implemented in quantum SDKs, each requiring 2 controlled-NOT gates. If we are able to define a set of encodings with properties not available to a single static encoding, transformations between them may be inserted into the quantum circuit. This is exactly the approach we will use to parallelize simulation with Trotterization.

\section{Combinatorial Coverings}\label{sec:coverings} % Something on what the task is!
Partitioning all of the degree 4 fermionic operators of an $M$ mode ESH across $q$ QPUs can be interpreted as a combinatorial covering problem.

Given a set $\mathcal{V}$ such that $|\mathcal{V}| = v$, and $k,t \in \mathbb{N}$ such that $v \geq k \geq t$, a $(v,k,t)$ \emph{covering} is a set of $k$-subsets of $\mathcal{V}$ such that all $t$-subsets of $\mathcal{V}$ are included in at least one $k$-subset.~\cite{HandbookCombinatorialDesigns2007} If the $k$-subsets can be combined such that they form parallel classes in which each member of $\mathcal{V}$ appears exactly once, the covering is said to be \emph{resolvable}. 

Significant effort has been devoted to finding the minimum number of $k$-subsets, the covering number $C(v,k,t)$, required for a given $(v,k,t)$.~\cite{HandbookCombinatorialDesigns2007,LaJollaCoveringRepository} Gordon et al.~\cite{NewConstructionsCovering1995} give a general formula for resolvable coverings which can be obtained from the $\beta$-flats of affine geometries $AG(\alpha,q)$:
\begin{equation}
  C(q^{\alpha}, q^{\beta}, \beta+1) \leq q^{\alpha-\beta}\begin{bmatrix} \alpha \\ \beta \end{bmatrix}_q
  \label{eq:covering-number}
\end{equation}
where $\begin{bmatrix} \alpha \\ \beta \end{bmatrix}_q$ is the q-binomial coefficient. Note that affine geometries can only be defined where $q$ is a prime power. In the case that $\beta=\alpha-1$, the above gives an equality. The $(16,8,4)$ covering derived from $AG(4,2)$ is given in Appendix~\ref{app:16-8-4}.

One final feature of covering designs we make use of is \emph{inflation}. By replacing each point in $\mathcal{V}$ by a set of size $n$, $v \to n v$ and increasing the size of $k$ in proportion $k \to nk$ we obtain a new design $(n v, n k, t)$. The covering number for this is upper bounded by the covering number of the uninflated design.~\cite{NewConstructionsCovering1995}

\begin{equation}
  C(n q^{\alpha}, n q^{\beta}, \beta+1) \leq q^{\alpha-\beta}\begin{bmatrix} \alpha \\ \beta \end{bmatrix}_q
\end{equation}

To parallelize a fermionic Hamiltonian with $M$ modes ($\mathcal{V} = [0,M)$) each fermionic operator ($t$-subset) must be fully contained in a QPU register ($k$-subset) at least once. Using a resolvable covering ensures all modes are included in each parallel class of $k$-subsets. Each QPU register can then be mapped to one of these $k$-subsets by an appropriate enumeration of the JW encoding.

To obtain an optimal resolvable covering of the degree 4 ESH with modes evenly split across $q$ QPUs, we set $\beta=3$ (and therefore $\alpha$=4) in Eq.~\eqref{eq:covering-number}. Inflating this design gives,

\begin{equation}
  C(n q^4, n q^{3}, 4) \le q \frac{1-q^4}{1-q}
\end{equation}

For a system of $M$ modes, we therefore require a register size of at least $M/q$ in each processor and $\mathcal{O}(q^3)$ encodings are needed, each implementing $q$ of the $k$-subsets in parallel. Equivalently, we may set a maximum register size $R$, and determine that a network of at most $R^{1/3}$ QPUs can parallelize a system of at most $R^{4/3}$ modes.

For systems in which $q^4 \nmid M$ we may pad the Hamiltonian, assigning zero coefficient to all terms containing a mode with index $\ge M$. These padding modes do not need to be parallelized. This will result in an unequal split of real terms and modes across QPUs, requiring more than $M/q$ qubits on each device. However, the padding required is upper bounded by $q^4 - 1$ and decreases as a proportion of $M$ for inflated designs.

\section{Parallel Fermionic Trotterization}\label{sec:pft}
The procedure to parallelize across two QPUs is described in Algorithm~\ref{alg:prs} and straightforwardly generalizes to larger networks. Required inputs are a Majorana Hamiltonian, and the indices of qubits included in two subsets, $A$ and $B$. The output is a set of pairs, each containing an encoding $\mathcal{E}^j$ and the set of fermionic Hamiltonian terms which $\mathcal{E}^j$ parallelizes, $h_j$. Each iteration updates the mode enumeration of the JW encoding, partitions terms according to their support as described in Section~\ref{sec:support} and updates the set of circuit slices $S$.

\begin{algorithm}[H]
  \caption{Parallelization by Dynamic Encoding}\label{alg:prs}
  \begin{algorithmic}[1]
    \Require {$H_\gamma$, $A$, $B$}
    \State $j \gets 0$
    \State $H^\mathrm{AB}_0 \gets H_\gamma$
    \State $S \gets \emptyset$
    \While{$|H^\mathrm{AB}_j| > 0$}
      \State $\mathcal{E}^{j} \gets \mathrm{update\_enumeration}(H^\mathrm{AB}_{j})$
      \State $H^A_j, H^B_j, H^\mathrm{AB}_{j+1} \gets$ partition($\mathcal{E}^j$, $H^\mathrm{AB}_{j}$, $A$, $B$)
      \State $h_j \gets H^A_j \cup H^B_j$
      \State $S \gets S \cup \{(\mathcal{E}^j, h_j)\}$
      \State $j \gets j+1$
    \EndWhile
    \State \Return $S$
  \end{algorithmic}
\end{algorithm}

The output encodings and Hamiltonian subsets in $S$ can be used to construct a Trotterized simulation circuit, with the whole-system Hamiltonian $H = \sum_j h_j$. Using Eq.~\eqref{eq:first-order-Trotter}, we construct a first order Trotter circuit as an example. This construction is shown diagrammatically in Fig.~\ref{fig:flat}.

Beginning with an initial encoding $\mathcal{E}^k$ for the whole circuit $U_f$,
\begin{equation}
  U_q = \mathcal{E}^k (U_f) \approx \mathcal{E}^k \Big(\prod_{j} e^{-i h_j t}\Big)
\end{equation}
The circuit is split into $N$ circuit slices, one for each $(\mathcal{E}, h)$ pair in $S$,
\begin{equation}
  U_q = \prod_j \mathcal{E}^k (e^{-i h_j t})
\end{equation}

Using Eq.~\eqref{eq:encoding-transform}, we map from $\mathcal{E}^k$ to $\mathcal{E}^j$ for each $h_j$,
\begin{equation}
  U_q = \prod_j V_{k j}\mathcal{E}^j (e^{-i h_j t})V_{k j}^{\dagger}
\end{equation}
Applying the encoding separates the circuit slice into parallel components. At this point the assignment of register labels $A$ and $B$ can be chosen to minimize communication cost.
\begin{equation}
  U_q = \prod_j V_{k j} (e^{-i H_j^A t} \otimes e^{-i H_j^B t}) V_{k j}^{\dagger}
\end{equation}

As we have only altered the encoding enumeration scheme, transformations between encodings $V_{ij}$ are composed of fSWAP networks. 
Using the result of Constantinides et al., swap networks can be implemented efficiently in depth~\cite{LowdepthFermionRouting2025}, though care must be taken not to unnecessarily increase communication cost. Note also that adjacent swap networks $V_{k i}^{\dagger}V_{k j}$ can be easily combined into a single permutation of modes.

\begin{figure}[t]
  \centering
  \newcommand{\circuitA}{%
  \begin{quantikz}[row sep=0.15cm, column sep=0.2cm, ampersand replacement=\&]
    \lstick{$\ket{0_q}$} \& \gate[3][\qcunit]{\mathcal{E}^k (e^{-i h_0 t} e^{-i h_1 t} e^{-i h_2 t})} \& \qw  \\
    \lstick{$\ket{\ldots}$} \& \qw \& \qw  \\
    \lstick{$\ket{M_q-1}$} \& \qw \& \qw 
  \end{quantikz}}
  \newcommand{\circuitB}{%
  \begin{quantikz}[row sep=0.15cm, column sep=0.2cm, ampersand replacement=\&]
    \lstick{$\ket{0_q}$} \& \gate[3][\qcunit]{\mathcal{E}^k (e^{-i h_0 t})} \& \gate[3][\qcunit]{\mathcal{E}^k (e^{-i h_1 t})} \& \gate[3][\qcunit]{\mathcal{E}^k (e^{-i h_2 t})} \& \qw  \\
    \lstick{$\ket{\ldots}$} \& \qw \& \qw \& \qw \& \qw  \\
    \lstick{$\ket{M_q-1}$} \& \qw \& \qw \& \qw \& \qw 
  \end{quantikz}}
  \newcommand{\circuitC}{%
  \begin{quantikz}[row sep=0.15cm, column sep=0.2cm, ampersand replacement=\&]
    \lstick{$\ket{0_q}$} \& \gate[3]{V_{k 0}} \& \gate[3][\qcunit]{\mathcal{E}^0(e^{-i h_0 t})} \& \gate[3]{V_{k 0}^\dagger} \& \gate[3]{V_{k 1}} \& \gate[3][\qcunit]{\mathcal{E}^1(e^{-i h_1 t})} \& \gate[3]{V_{k 1}^\dagger} \& \gate[3]{V_{k 2}} \& \gate[3][\qcunit]{\mathcal{E}^2 (e^{-i h_2 t})} \& \gate[3]{V_{k 2}^\dagger} \& \qw  \\
    \lstick{$\ket{\ldots}$} \& \qw \& \qw \& \qw \& \qw \& \qw \& \qw \& \qw \& \qw \& \qw \& \qw  \\
    \lstick{$\ket{M_q-1}$} \& \qw \& \qw \& \qw \& \qw \& \qw \& \qw \& \qw \& \qw \& \qw \& \qw 
  \end{quantikz}}
  \newcommand{\circuitD}{%
  \begin{quantikz}[row sep=0.15cm, column sep=0.2cm, ampersand replacement=\&]
    \lstick{$\ket{0_q}$} \& \gate[4]{V_{k 0}} \& \gate[2][\qcunit]{e^{-i H^A_0 t}} \& \gate[4]{V_{k 0}^\dagger} \& \gate[4]{V_{k 1}} \& \gate[2][\qcunit]{e^{-i H^A_1 t}} \& \gate[4]{V_{k 1}^\dagger} \& \gate[4]{V_{k 2}} \& \gate[2][\qcunit]{e^{-i H^A_2 t}} \& \gate[4]{V_{k 2}^\dagger} \& \qw  \\
    \lstick{$\ket{\ldots}$} \& \qw \& \qw \& \qw \& \qw \& \qw \& \qw \& \qw \& \qw \& \qw \& \qw  \\
    \lstick{$\ket{\ldots}$} \& \qw \& \gate[2][\qcunit]{e^{-i H^B_{0} t}} \& \qw \& \qw \& \gate[2][\qcunit]{e^{-i H^B_{1} t}} \& \qw \& \qw \& \gate[2][\qcunit]{e^{-i H^B_{2} t}} \& \qw \& \qw  \\
    \lstick{$\ket{M_q-1}$} \& \qw \& \qw \& \qw \& \qw \& \qw \& \qw \& \qw \& \qw \& \qw \& \qw 
  \end{quantikz}}
  \fitcircuit{1}{\circuitA}

  \vspace{1ex}
  \fitcircuit{3}{\circuitB}

  \vspace{1ex}
  \fitcircuit{3}{\circuitC}

  \vspace{1ex}
  \fitcircuit{3}{\circuitD}
  \caption{Qubit circuit for parallelized first order Trotterization with 3 circuit slices.}
  \label{fig:flat}
\end{figure}

\subsection{Encoding Enumeration Update}\label{sec:encoding-optimisation}

Several approaches can be taken to Line 5 of Algorithm~\ref{alg:prs}. The covering designs we have presented in Section~\ref{sec:coverings} provide a set of encoding enumerations which are guaranteed to fully parallelize the Hamiltonian. We may take another approach, selecting a new encoding during each iteration.

Firstly, given enough iterations the procedure will terminate if enumerations of JW are selected randomly. Of course this becomes infeasible with increasing system size as the number of even partitions is $\frac{1}{2} \binom{M}{M/2}$. Very often enumerations will be chosen which do not parallelize any terms, in which case their associated circuit slice is just the identity.

By using an optimization method which algorithmically reduces the length of encoded Pauli-strings, such as TOPP-HATT, significantly fewer iterations may be needed than selecting random enumerations.~\cite{OptimisedFermionQubitEncodings2026} As terms are removed from the Hamiltonian with each iteration, the enumeration can be re-optimized to give the shortest possible Pauli-strings for the remaining terms. By doing so, fewer terms will cross the cuts between registers with each iteration.

Describing the Hamiltonian as a hypergraph of $M$ nodes, with each non-zero Hamiltonian term giving a hyperedge, we can consider the task of parallelizing terms as finding a q-partition of the hypergraph. A general purpose hypergraph partitioning algorithm such as KaHyPar can be used.~\cite{HighQualityHypergraphPartitioning2021} The output enumeration must then place the modes in each partition as contiguous blocks of the JW encoding.

In the following, we present results using each of these on-the-fly methods as comparisons to the covering design approach.

\section{Communication Cost}\label{sec:communication-cost}
Our main result is an upper bound on communication cost of parallelization which scales linearly in the number of fermionic modes $M$ and the Trotter number $r$, at the cost of third power scaling in the number of QPUs $q$, giving $\mathcal{O}(M q^3 r)$. 

The construction of Algorithm~\ref{alg:prs} allows us to determine bounds on the number of inter-QPU fSWAP gates, and therefore the communication cost between QPUs. In the worst case, all $M$ modes are transmitted between processors at each iteration. For $N$ iterations, the total number of swaps for first order Trotterization then scales with $\mathcal{O}(M N)$. This increases to $\mathcal{O}(M N r)$ with Trotter number $r$. Using an affine geometry derived covering design to define encodings results in $N$ independent of $M$, instead scaling with $q^3$.

\begin{figure}[h!]
  \centering
  \includegraphics[width=\columnwidth]{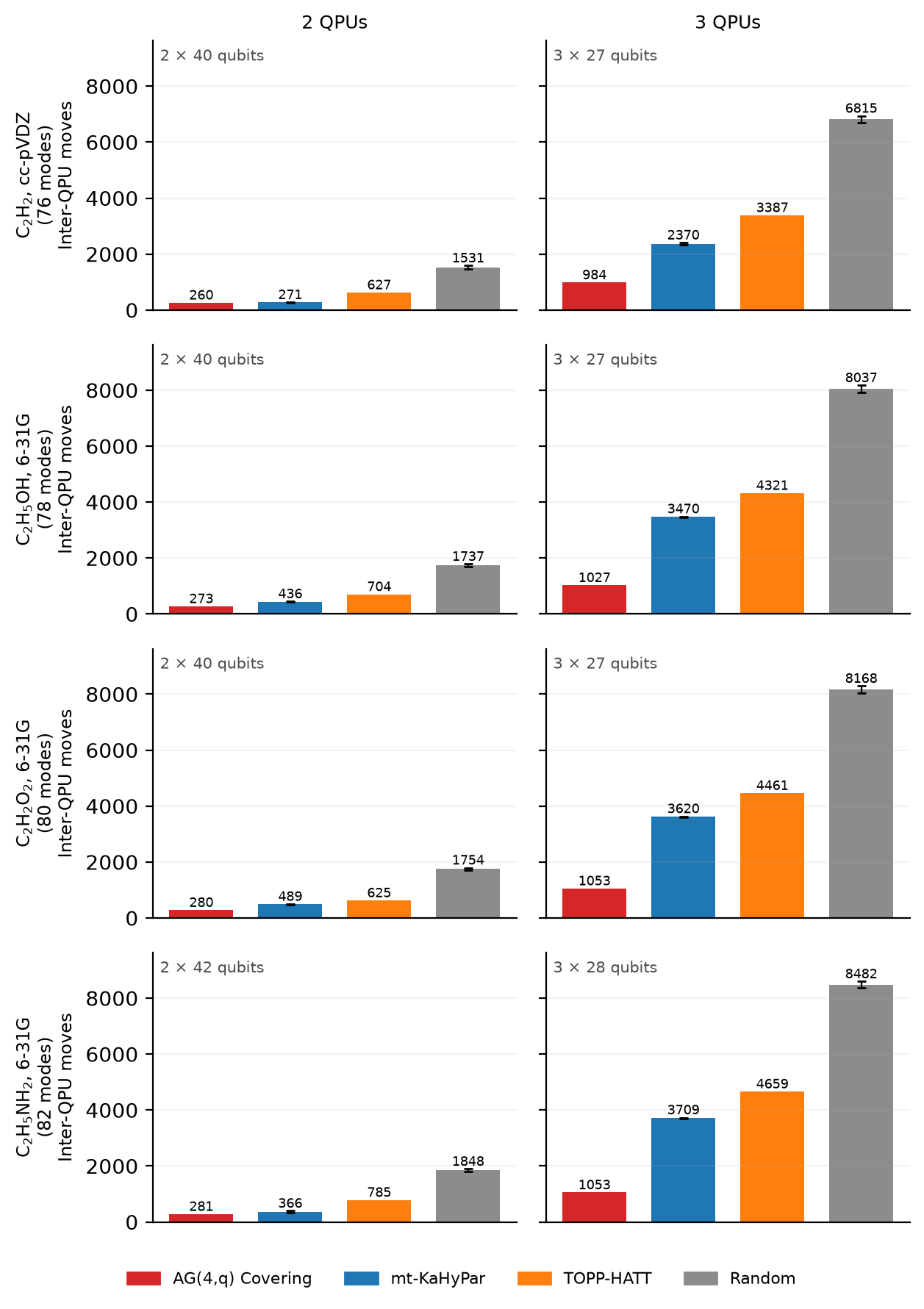}
  \caption{Communication cost of a single Trotter step for $C_2 H_2$ (cc-pVDZ), $C_2 H_5 OH$ (6-31G), $C_2 H_2 O_2$ (6-31G) and $C_2 H_5 NH_2$ (6-31G). Bars in grey are random enumerations, orange are TOPP-HATT optimized, blue are hypergraph partitions using mt-KaHyPar and in red are $AG(4,q)$ covering results. Numerical values are given in Table~\ref{tab:qpu-comm}.}
  \label{fig:iterations}
\end{figure}

We demonstrate our method with a set of molecular Hamiltonians, each containing approximately 80 fermionic modes. This number was chosen as designs for both two ($80=5 \times 2^4$) and three ($81=3^4$) QPUs can be compared. Figure~\ref{fig:iterations} shows the required communication cost in terms of fSWAP gates for a single Trotter step of $C_2 H_2$ (cc-pVDZ), $C_2 H_5 OH$ (6-31G), $C_2 H_2 O_2$ (6-31G) and $C_2 H_5 NH_2$ (6-31G). We compare the communication cost of parallelization with $AG(4,q)$ coverings to random JW enumerations, TOPP-HATT optimization and hypergraph partitioning with mt-KaHyPar.

As described in Section~\ref{sec:encoding-optimisation}, when using the covering method Hamiltonians are padded if required. In the case of $C_2 H_5 NH_2$, padding of 14 modes is needed for $q=2$ and 80 modes for $q=3$. However, after deleting padded modes from the resulting covering design, required register sizes are only $42$ and $28$ respectively. Alternative methods also make use of these increased register sizes.

Molecular geometries were obtained from PubChem. \cite{PubChem2021New2021} Hamiltonian coefficients were calculated using PySCF and OpenFermion. \cite{PythonbasedSimulationsChemistry2017, OpenfermionElectronicStructure2017} Encodings were defined and applied with ferrmion. \cite{UCLCCSFerrmionV0312025}

% \section{Trotter Error}\label{sec:trotter-error}
% Thus far we have presented a method for parallelization with communication cost in term of the Trotter number. We now consider the impact of this approach on the required Trotter number and the communication cost for a given simulation accuracy.

% For a Hamiltonian $H = \sum_{j} H_j$, the Trotter number required for a $p$-th order Trotterization to error $\epsilon$ is given by\cite{TheoryTrotterError2021}:

% \begin{equation}
%   r = \mathcal{O}\big(\frac{\alpha^{1/p} t^{1+1/p}}{\epsilon^{1/p}} \big)
% \end{equation}

% where $alpha$ is a sum over the spectral norms of all permuatatons of nested commutators of terms $H_j$. 

% \begin{equation}
%   \alpha = \sum_{j_{p}} \dots \sum_{j_0} ||[H_{j_p}, \dots [H_{j_1}, H_{j_0}] \dots ]||
% \end{equation}

\section{Discussion}\label{sec:discussion}
Our method shows an improvement over a previous complexity bound describing distributed Trotterization of Pauli encoded Hamiltonians given by Feng et al.~\cite{DistributedQuantumSimulation2024a}, which applies when a static encoding is used for the fermionic ESH. Their analysis gives communication cost scaling $\mathcal{O}(|H^\mathrm{AB\dots}| q r)$ where $|H^\mathrm{AB\dots}| \propto M^4$ is the number of Hamiltonian terms with supports spanning qubits in multiple QPUs. 

A recent publication presented bounds on two-QPU distribution complexity of the double-factorized (DF) electronic structure Hamiltonian.~\cite{DistributionComplexityElectronic2026} With this method, the inter-QPU communication cost of Trotterization was found to scale linearly in both the number of fermionic modes $M$ and rank of the factorization $L$, resulting in $\mathcal{O}(M L r)$ scaling. Exact results are found for $L = M(M+1)/2$. Empirical studies suggest chemical accuracy may be reached with $L$ sub-linear in $M$, though this may be dependent on the target Hamiltonian. This dependence of $L$ on $M$ guarantees it cannot meet the $\mathcal{O}(M r)$ scaling we have demonstrated. One point of distinction between their approach and our own is that we work directly with the ESH, without applying double-factorization. To obtain the reduction in scaling of Trotterization circuit depth from $\mathcal{O}(M^4)$ to $\mathcal{O}(M^3)$ which DF provides, each circuit slice could be transformed to DF form, but must result in a state with the same correspondence of mode-occupation and spin as that slice's JW enumeration. Similarly, each slice may be transformed to the plane-wave basis using a fermionic Fourier transform~\cite{QuantumSimulationElectronic2018}, or optimized using Majorana-SWAP networks~\cite{ImprovingFermionicVariational2025}.

We do not address the effect of circuit slice ordering on communication cost or Trotter error.~\cite{DistributedQuantumSimulation2024a} Necaise et al. do suggest a pathfinding approach to order slices such that communication cost is minimized.~\cite{DistributionComplexityElectronic2026} One further consideration which we leave open is the effect of overlap in the sets of parallelizable terms of encodings used. Our method removes parallelized Hamiltonian terms with each iteration, leaving open the possibility that some terms would be parallelized by more than one encoding. This is certainly the case, and clearly worth considering, if using a covering design which covers terms more than once. Assuming all parallelized slices are implemented without error, Trotter error resulting from partitioning the Hamiltonian into slices still remains. This may be minimized both by ordering slices, and selecting which slice terms are assigned to. We leave this open for future work.

The JW based Accordion scheduler exploits properties of the JW encoding to reduce circuit metrics. Accordion's grouping and parallelization procedure targets a single QPU.~\cite{LinearComplexityFermionic2026} In our method, encoding enumerations are used to define an efficient partition between QPUs, with fSWAP gates acting between groups in the manner of Accordion's inter-group scheduler. Our method could also be used to parallelize circuits on a single device. Evaluating this in conjunction with other optimizations in Accordion would be an interesting extension of this work. Given Accordion imposes a linear qubit connectivity to target devices, it could also be employed across multiple QPUs if a suitable qubit assignment can be found. However, it does not have a mechanism to prioritize inter-QPU entangling operations, which are significantly more costly than those performed on-device.~\cite{ReviewDistributedQuantum2025} Our method could be used to parallelize, with Accordion applied separately on each QPU.

Another recently published framework, COSMA, employs a genetic algorithm to optimize communication cost of fermionic simulation using ternary tree encodings. Optimization is performed over the summed communication costs of each individual Pauli operator.~\cite{COSMACommunicationawareOptimization2026} This process results in a static fermion-qubit encoding which is applied to all Hamiltonian terms. Using an evolutionary algorithm with a fixed limit of generations and population size results in decreasing performance as the system size increases. In contrast, the communication cost of our method is fixed for a given Hamiltonian and number of QPUs.

Within the broader distributed quantum computing literature, our approach resembles the \emph{Fine Grained Partitioning} (FGP) methods. By exploiting clustering of gates around a subset of qubits, more efficient qubit routings can be found to minimize communication cost for a distributed circuit \cite{ReviewDistributedQuantum2025}. This results in parallel circuit slices interspersed with inter-QPU communication operations, in the same manner as our method. However, rather than beginning from a circuit to which heuristics are applied, we begin with a partition and define the sequence of gates from this. Inverting the method in this way is only possible as the ordering of terms in a Trotter decomposition is not fixed.

\section{Conclusion}
We have demonstrated a method to efficiently parallelize Trotterization of fermionic Hamiltonians. Communication cost is dependent on the number of modes $M$ of the Hamiltonian, the Trotter number $r$, and number of circuit slices produced $N$, with scaling upper bounded by $\mathcal{O}(M N r)$. Using an affine geometry derived covering design gives $N\propto q^3$ for $q$ QPUs, and therefore communication cost $\mathcal{O}(M q^3 r)$.
Output circuits are given in terms of JW encoded fermionic Hamiltonians, which can be transformed to other encodings and simulated using any method which may be applied to the whole Hamiltonian.

Use of this method in combination with existing circuit compilation and distribution methods may provide further benefit to real use cases, including on-device parallelism.
Future investigation should also seek to exploit degrees of freedom in this method to optimize communication cost and Trotter error.

\section*{Acknowledgements}
M.W.d.l.B. acknowledges support from the Engineering and Physical Sciences Research Council (EPSRC) (Grant No. EP/S021582/1). 
P.V.C. acknowledges support from the UK Engineering and Physical Sciences Research Council (EPSRC) under the project “Software Environment for Actionable \& VVUQ-evaluated Exascale Applications (SEAVEA)” (Grant No. EP/W007711/1).
The authors would also like to thank Evan Sutcliffe and Tom Bickley for helpful questions on a previous version of this manuscript.

\section*{AI Usage Disclosure}
This work was motivated by and develops previous work of the authors. Claude Opus 5.5 was used to prototype code, generate figures, search literature and proof-read for minor typographical and grammatical errors. The authors retain full responsibility for any errors or omissions.

\bibliography{references}

% ---------------------------------------------------------------------------
% Appendix (from appendix.typ)
% ---------------------------------------------------------------------------
\clearpage
\onecolumngrid
\appendix

\section{Affine Geometries}\label{app:16-8-4}

Given an $\alpha$-dimensional vector space $W$ over the finite field $GF(q)$, the set of all cosets of subspaces of $W$ define the affine geometry $AG(\alpha,q)$. Points, or 0-flats, of the affine geometry $AG(\alpha,q)$, are given by cosets of the zero vector. In general, the i-flats are given by cosets of i-dimensional subspaces and contain $q^i$ points.

A $(16,8,4)$ covering can be defined from the points and 3-flats of $AG(4,2)$.~\cite{DesignTheory1999,NewConstructionsCovering1995} Table \ref{tab:covering} shows the 30 size 8 subsets arranged into 15 pairs.

\begin{table}[h!]
\begin{tabular}{|c|c|c|}
  \hline
  $i$ & $A_i$ & $B_i$ \\
  \hline
  0 & 0, 1, 2, 3, 4, 5, 6, 7 & 8, 9, 10, 11, 12, 13, 14, 15 \\
  1 & 0, 1, 2, 3, 8, 9, 10, 11 & 4, 5, 6, 7, 12, 13, 14, 15 \\
  2 & 0, 2, 4, 6, 8, 10, 12, 14 & 1, 3, 5, 7, 9, 11, 13, 15 \\
  3 & 0, 1, 4, 5, 8, 9, 12, 13 & 2, 3, 6, 7, 10, 11, 14, 15 \\
  4 & 0, 1, 2, 3, 12, 13, 14, 15 & 4, 5, 6, 7, 8, 9, 10, 11 \\
  5 & 0, 2, 4, 6, 9, 11, 13, 15 & 1, 3, 5, 7, 8, 10, 12, 14 \\
  6 & 0, 1, 4, 5, 10, 11, 14, 15 & 2, 3, 6, 7, 8, 9, 12, 13 \\
  7 & 0, 2, 5, 7, 8, 10, 13, 15 & 1, 3, 4, 6, 9, 11, 12, 14 \\
  8 & 0, 1, 6, 7, 8, 9, 14, 15 & 2, 3, 4, 5, 10, 11, 12, 13 \\
  9 & 0, 3, 4, 7, 8, 11, 12, 15 & 1, 2, 5, 6, 9, 10, 13, 14 \\
  10 & 0, 2, 5, 7, 9, 11, 12, 14 & 1, 3, 4, 6, 8, 10, 13, 15 \\
  11 & 0, 1, 6, 7, 10, 11, 12, 13 & 2, 3, 4, 5, 8, 9, 14, 15 \\
  12 & 0, 3, 4, 7, 9, 10, 13, 14 & 1, 2, 5, 6, 8, 11, 12, 15 \\
  13 & 0, 3, 5, 6, 8, 11, 13, 14 & 1, 2, 4, 7, 9, 10, 12, 15 \\
  14 & 0, 3, 5, 6, 9, 10, 12, 15 & 1, 2, 4, 7, 8, 11, 13, 14 \\
  \hline
\end{tabular}
\caption{Subsets of the $(16,8,4)$ covering design derived from $AG(4,2)$.}
\label{tab:covering}
\end{table}

\section{Numerical Results}
\begin{table*}[h!]
  \caption{Communication cost $C$ for a single Trotter step of molecular Hamiltonians, given in terms of fSWAP gates. $M$ gives the number of fermionic modes. $Register$ gives the number and size of QPU registers in the network. Results relate to parallelization with an affine geometry $AG(4,q)$ derived covering, mt-KaHyPar, TOPP-HATT and random enumerations. For mt-KaHyPar and random results, means of 10 runs are given with standard deviations in parentheses.}
  \label{tab:qpu-comm}
  \begin{ruledtabular}
  \begin{tabular}{llrcrrrr}
    Molecule & Basis & $M$ & Register & $C_\mathrm{covering}$ & $C_\mathrm{mt\text{-}K}$ & $C_\mathrm{T\text{-}H}$ & $C_\mathrm{random}$ \\
    \hline
    \multicolumn{8}{l}{\emph{2 QPUs}} \\
    C$_2$H$_2$ & cc-pVDZ & 76 & $2 \times 40$ & 260 & 271.3 (4.6) & 627 & 1530.5 (54.7) \\
    C$_2$H$_5$OH & 6-31G & 78 & $2 \times 40$ & 273 & 436.1 (13.9) & 704 & 1736.5 (47.1) \\
    C$_2$H$_2$O$_2$ & 6-31G & 80 & $2 \times 40$ & 280 & 488.9 (7.4) & 625 & 1754.1 (39.5) \\
    C$_2$H$_5$NH$_2$ & 6-31G & 82 & $2 \times 42$ & 281 & 365.5 (28.7) & 785 & 1848.5 (42.3) \\
    \hline
    \multicolumn{8}{l}{\emph{3 QPUs}} \\
    C$_2$H$_2$ & cc-pVDZ & 76 & $3 \times 27$ & 984 & 2369.8 (36.2) & 3387 & 6814.9 (119.2) \\
    C$_2$H$_5$OH & 6-31G & 78 & $3 \times 27$ & 1027 & 3470.3 (14.6) & 4321 & 8037.3 (126.7) \\
    C$_2$H$_2$O$_2$ & 6-31G & 80 & $3 \times 27$ & 1053 & 3619.5 (12.9) & 4461 & 8168.5 (114.3) \\
    C$_2$H$_5$NH$_2$ & 6-31G & 82 & $3 \times 28$ & 1053 & 3709.0 (20.9) & 4659 & 8482.5 (104.3) \\
  \end{tabular}
  \end{ruledtabular}
\end{table*}

\end{document}